\documentclass[12pt, journal, letterpaper]{IEEEtran}
\usepackage{amsmath}
\usepackage{graphicx,subfigure,comment}
\usepackage[table]{xcolor}
\usepackage{url} 
\usepackage{dsfont}
\usepackage{cite}
\usepackage{tabularx}
\usepackage{soul}
\usepackage{todonotes}
\usepackage{balance}

\newcommand{\Fig}[1]{Fig.~\ref{fig:#1}}

\newcommand{\Sec}[1]{Sec.~\ref{sec:#1}}

\begin{document}

\title{Choose, not Hoard: Information-to-Model Matching for Artificial Intelligence in O-RAN}
\author{\IEEEauthorblockN{Jorge Martín-Pérez}\IEEEauthorrefmark{1}, \IEEEauthorblockN{Nuria Molner}\IEEEauthorrefmark{2}\IEEEauthorrefmark{4}, \IEEEauthorblockN{Francesco Malandrino}\IEEEauthorrefmark{3}, \IEEEauthorblockN{Carlos~Jesús~Bernardos}\IEEEauthorrefmark{1}, \IEEEauthorblockN{Antonio de la Oliva}\IEEEauthorrefmark{1}, \IEEEauthorblockN{David Gomez-Barquero}\IEEEauthorrefmark{2}\\
\IEEEauthorblockA{\IEEEauthorrefmark{2} iTEAM Research Institute - Universitat Polit\`ecnica de Val\`encia, Spain} \\
\IEEEauthorblockA{\IEEEauthorrefmark{1} Universidad Carlos III de Madrid, Spain}
\IEEEauthorblockA{\IEEEauthorrefmark{3} CNR-IEIIT, Torino, Italy\vspace*{-1cm}}
\thanks{\IEEEauthorrefmark{4}corresponding author: numolsiu@iteam.upv.es \\ 
	\copyright 2022 IEEE. Personal use of this material is permitted. Permission from IEEE must be obtained for all other uses, in any current or future media, including reprinting/republishing this material for advertising or promotional purposes, creating new collective works, for resale or redistribution to servers or lists, or reuse of any copyrighted component of this work in other works.}
}
\maketitle

\begin{abstract}

Open Radio Access Network (O-RAN) is an emerging paradigm, whereby virtualized network infrastructure elements from different vendors communicate via open, standardized interfaces. A key element therein is the RAN Intelligent Controller (RIC), an Artificial Intelligence (AI)-based controller.
Traditionally, all data available in the network has been used to train a single AI model to be used at the RIC. This paper introduces, discusses, and evaluates
the creation of {\em multiple} AI model instances at different RICs,
leveraging information from some (or all) locations for their training.
This brings about a flexible relationship between gNBs, the AI models used to control them, and the data such models are trained with.
Experiments with real-world traces show 
how using \emph{multiple} AI model instances
that \emph{choose}
training data from specific locations improve the
performance of traditional approaches following the {\em hoarding} strategy.
\end{abstract}

\section{Introduction}
\label{sec:intro}

Virtual Radio Access Network (vRAN) is arguably one of the most exciting recent innovations of the networking ecosystem. It is enabled by the Software-Defined Networking (SDN) approach, and allows the functions traditionally performed by base stations (currently gNBs) to be {\em virtualized} and {\em split} across multiple network nodes, including newly-introduced entities called Central Units (CUs), Distributed Units (DUs), and Radio Units (RUs). Such a functional split allows different decisions to be made at different nodes {\em and} with different time scales.
For example, RUs can perform real-time radio management, while CUs can adjust higher-level resource allocation at longer time scales. The different CU, DU, RU units corresponding to different gNBs can now be hosted in edge or cloud servers, sharing location in some cases and reducing costs for the operators through the remote management of the components thanks to its virtualized nature.

The promising results of vRAN gave rise to initiatives, such as Open Radio Access Network (O-RAN) or Cisco's Open vRAN Ecosystem Group, aiming at creating an open and interoperable RAN ecosystem where open APIs and interfaces can be integrated connecting different vendors components. O-RAN~\cite{garcia2021ran} has been so far the vRAN initiative receiving more attention, also thanks to the open-source community created around it.
 
In addition to the vRAN components, O-RAN introduces a new element called RAN Intelligent Controller (RIC), implementing arbitrary resource allocation and management policies via closed-control loops. Different RICs can run at different time scales, e.g., near-real-time (with latencies of less than 10~ms) and non-real-time (with latencies of several seconds). Owing to their (relatively) relaxed time requirements, non-real-time RICs can leverage Artificial Intelligence (AI) and Machine Learning (ML) for their decisions. RICs can collect from DUs information about the current state of the network, process such data, and instruct RUs accordingly~\cite{bonati2021intelligence}.
The importance of AI in O-RAN is such that a
dedicated working group~\cite{ORAN_WG2_AI_ML}
has been created to define use cases and specify which
components should host the AI/ML-based intelligence.

\begin{figure}[t]
    \centering
    \includegraphics[width=\columnwidth]{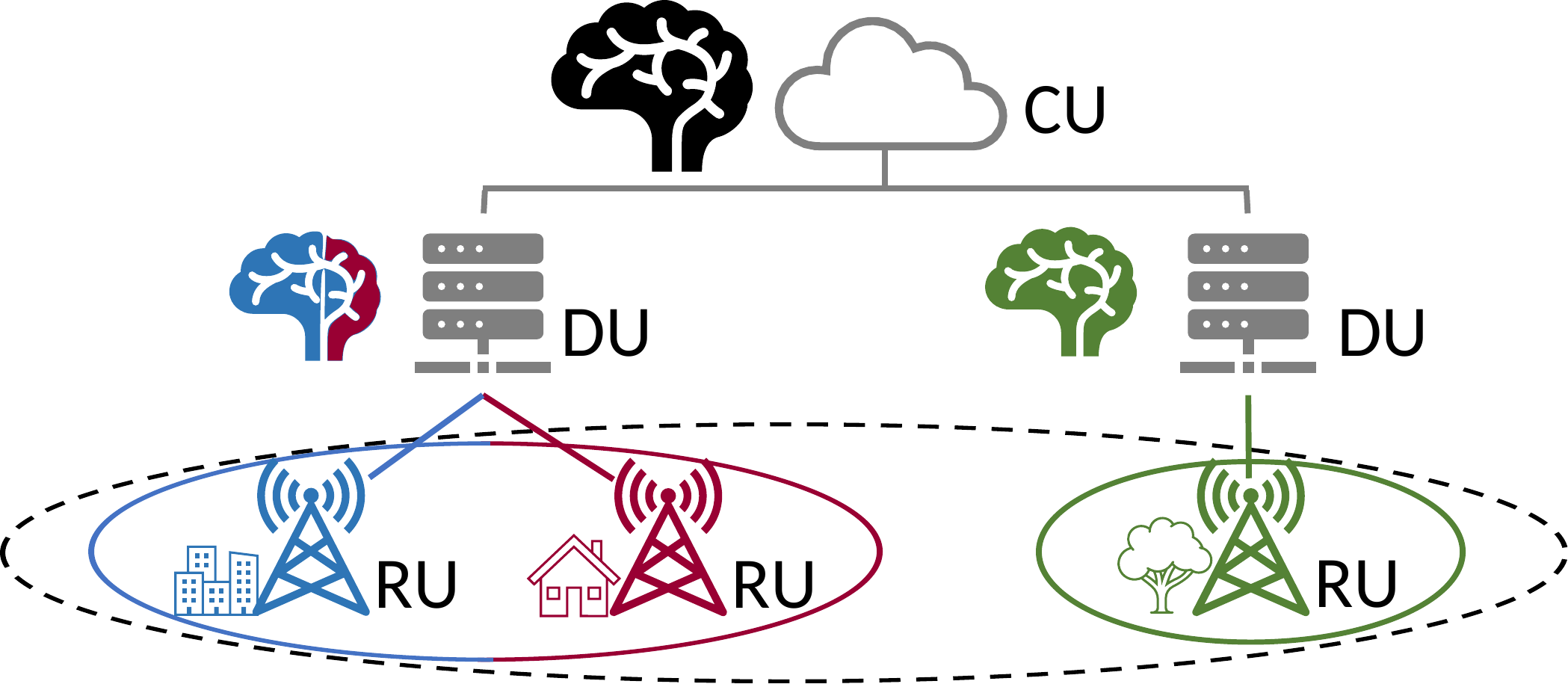}
    \caption{
        Choosing (solid lines) vs hoarding (dashed lines) data
         in a scenario with urban (blue),
        residential (purple) and rural (green)
        locations.
        Choosing results in specific model
        instances for urban+residential
        (blue/purple brain)
        and rural (green brain) locations.
        Hording results in a global
        model instance (black brain).
        \label{fig:archi_combined}
    }
\end{figure}

AI/ML techniques currently tailored to O-RAN scenarios are the subject of a vast body of research, as detailed in \Sec{existing}, with the majority of works being predicated on the notion that effective AI training requires \emph{hoarding} all existing data from all the sources (black brain in~\Fig{archi_combined}).
However, there are several reasons why this may not always be the best approach.
First, transferring data from all RUs to the RIC may incur long delays, hence, decisions may be based upon outdated information.
Furthermore, more data might result in a minor improvement in learning accuracy, at the price of significant longer training times. Finally, training an AI with data from unrelated locations (e.g., rural and urban areas) may even hurt the performance, unless extremely long training times are accepted~\cite{malandrino2021network}.

In this paper, we introduce, discuss, and evaluate the benefits of
\emph{choosing} which data is used to train an AI model in O-RAN.
Specifically, we propose to create \emph{multiple} model instances
running at different RICs, and to train them by {\em choosing} the most appropriate data,
even if it does not come from locations
under their control,
as exemplified in~\Fig{archi_combined}.
By doing so, we reap the twofold benefit of
(i) getting {\em better} learning,
as only relevant information is used, and
(ii) get {\em faster} learning, as having to move less information across
the network results in shorter network delays
and, usually, in
cost savings.

The reminder of this paper is organized as follows. In \Sec{existing}, we discuss state-of-the-art AI/ML in O-RAN scenarios, highlighting how most approaches seek to leverage all existing data to train a single AI model. ~\Sec{peva} uses a real-world scenario to compare
the~\emph{choosing} and~\emph{hoarding} approaches
in AI training.
Motivated by our experimental findings,
in~\Sec{omatch} we discuss in detail how the choosing approach tackles many of the major issues of AI in O-RAN scenarios.
Finally, \Sec{discussion} discusses open research issues, and \Sec{conclusion} concludes the paper.

\section{Existing AI/ML solutions for O-RAN}
\label{sec:existing}

As mentioned in \Sec{intro}, one of the core principles of O-RAN is to make networks intelligent and self-manageable~\cite{garcia2021ran,shafin2020artificial}. AI/ML is one of the key enabling technologies, with popular approaches including Reinforcement Learning (RL)~\cite{pamuklu2021reinforcement}, Deep Learning (DL)~\cite{ayala2019vrain}, and Federated Learning (FL)~\cite{cao2021user}.

One common application of AI in vRAN networks is to improve the usage of computational and networking resources, as done in~\cite{nikbakht2020unsupervised} and~\cite{pradhan2020computation}, which relies for this purpose on unsupervised and supervised DL techniques, respectively. Other works, use intelligence-powered optimization techniques for the semi-automated management of cellular networks integrated in real testbed environments~\cite{bonati2020cellos}. 

In the O-RAN WG2~\cite{ORAN_WG2_AI_ML}, AI is leveraged for traffic steering to trigger handovers to neighboring cells that are predicted to provide better performance to the terminal.
Bounded processing latency is a significant problem when deciding how and where to apply intelligence.
Swift decision making by RICs
received significant attention; for example,
\cite{bonati2021intelligence}~trains and validates models offline in the non real-time RIC, to then deploy them in the near real-time RIC to perform online decisions. The choice of the network node where the training and inference happen has a significant impact on training times and network delays, hence, this decision is critical. 

A related problem concerns how to learn from data located in different nodes, for which there are two different approaches, namely, centralized and distributed. Centralized learning requires to train one single model at one single server, that can be either in the edge or in the cloud. In this case, all the data is gathered at the server where the model is trained. This approach is considered in works such as~\cite{pradhan2020computation},~\cite{MLedge_TMC2021},~\cite{kazemifard2021minimum} and~\cite{bashir2019optimal}, which use AI to make network management decisions aimed at reducing end-to-end latency. The opposite approach is to train one model in a distributed fashion, at multiple cloud and edge servers. In this case, each server performs one epoch of the learning process with local data and exchanges the partial results with the rest of the training servers to include this information in the subsequent individual epochs, as in FL~\cite{cao2021user}. FL aims at creating a single, global model by averaging the local models of the different learning nodes. The advantage of this training is that, network latency tends to be lower as data is collected from close-by sources; furthermore, distributed approaches tend to preserve the privacy of the data. Works such as~\cite{wang2021network} and~\cite{malandrino2021network} follow this distributed approach.

Considering all the analyzed literature, we can observe a strong tendency to use {\em all} available data to obtain {\em one} single generic model -- trained in either a centralized or a distributed manner. The option of creating \emph{multiple} instances of the model, that can fit data of different nature, is as of yet unexplored.
Accordingly, \Sec{peva} leverages real-world cellular traces to verify our intuition that creating multiple model instances and {\em choosing} the data they are trained upon may beat the traditional approach of {\em hoarding} all data to train a single model instance.

\section{Choosing and hoarding\\in a real-world~scenario}
\label{sec:peva}

This section evaluates the effect of creating multiple AI models flexibly {\em choosing} their training data.
Specifically, we compare the extreme approaches of (i)
training multiple
model instances, each controlling one RU, and training them by {\em choosing}
data from a single RU (e.g., green brain in~\Fig{archi_combined}),
and (ii) \emph{hoarding} training data from all
RUs (e.g., black brain in~\Fig{archi_combined}) and training a single model instance.
More balanced approaches combining the two strategies may work better in practice; however, our main objective in this work is to establish if the {\em choosing} approach can yield a performance comparable to {\em hoarding}.

In order to draw realistic conclusions, we leverage two different real-world datasets.
The first
dataset~\cite{bonati2021intelligence} describes a 5G
network of 4~RUs at an urban scenario in Rome;
with each RU exchanging a fixed
amount of traffic
with 40~UEs that belong to
enhanced Mobile Broadband (eMBB),
Machine Type
Communication (MTC), and
Ultra Reliable Low Latency
Communication (URLLC) slices.
The second dataset\footnote{\url{https://challenge.aiforgood.itu.int/}}
describes a real-world LTE deployment of 3~RUs in three
residential areas of Barcelona: Les Corts, El Born,
and Poble Sec. 

It is important to highlight that the Rome and Barcelona datasets have very different levels of heterogeneity. The Rome one considers every RU receiving the same amount of traffic at every slice, while traffic levels are vastly different in the Barcelona one, as the traffic exchanged by each RU depends on how crowded their coverage area is.

For the Rome dataset~\cite{bonati2021intelligence},
the RIC's goal is to predict the performance, more specifically, the downlink (DL) bitrate experienced by
each user.
In line with~\cite{bonati2021intelligence}, we
use a feed-forward (FF) neural network (NN) with
2~hidden layers of 30~neurons with sigmoid
activations, inferring the
DL bitrate of each user equipment (UE) using as input: the network
slice, Modulation Coding Scheme (MCS), granted Physical Resource Blocks (PRBs), and buffer size.

For the Barcelona dataset,
the RIC's goal is to predict the aggregated DL
bitrate at each RU.
Following the lead of~\cite{ayala2019vrain},
to better adapt to the features of the trace,
we implement an encoder-decoder NN with
4~hidden layers having 16, 64, 32, and 32 neurons
with $\tanh$ activations. To infer the aggregated DL, the
NN is fed with information about the
MCS, PRBs, and number of Radio Network Temporary Identifier (RNTIs).

For the sake of simplicity, in all our experiments we train the NNs from scratch, i.e., with randomly-initialized weights. In practical scenarios, it is more common to start from partially-trained networks, e.g., under the {\em active learning} paradigm. In both cases the qualitative behavior is the same.

We train the FF and encoder-decoder 
NNs 
using the Adam optimizer with learning rates
of $10^{-6}$ and $10^{-5}$, respectively.
Data is transformed using $L_2$ normalization
for the FF NN, and MinMax($[-1,1]$)
normalization for the encoder-decoder NN.
The goal of both trainings is to minimize
the Mean Absolute Percentage Error (MAPE) for the validation set (20\% of the data).
Both NNs are implemented using PyTorch 1.10.0+cu102 on an Intel Xeon CPU E5-2670 @2.60GHz.


Within each scenario, we select one RU as our {\em target}; specifically, RU\textsubscript{4} in the Rome dataset,
and Poble~Sec RU in the Barcelona dataset.
We compare the performance of different setups:
a
\emph{hoarding} setup, leveraging data from all RUs, and multiple {\em choose} setups, each exploiting data coming from a single RU.
In all setups, we are chiefly interested in the trade-off betwen learning quality and the main factors limiting it, that is, time and data availability. Specifically,
we assess the performance resulting from
changing the quantity of training data
(\Sec{more-less-data}) and maximum
training time (\Sec{training-more-less-time}).
We further check how learning quality translates into the performance of a specific application, a
quality predictor xApp
(\Sec{quality-predictor}).

\subsection{Impact of the quantity of training data}
\label{sec:more-less-data}

\begin{figure}[t]
    \includegraphics[width=\columnwidth]{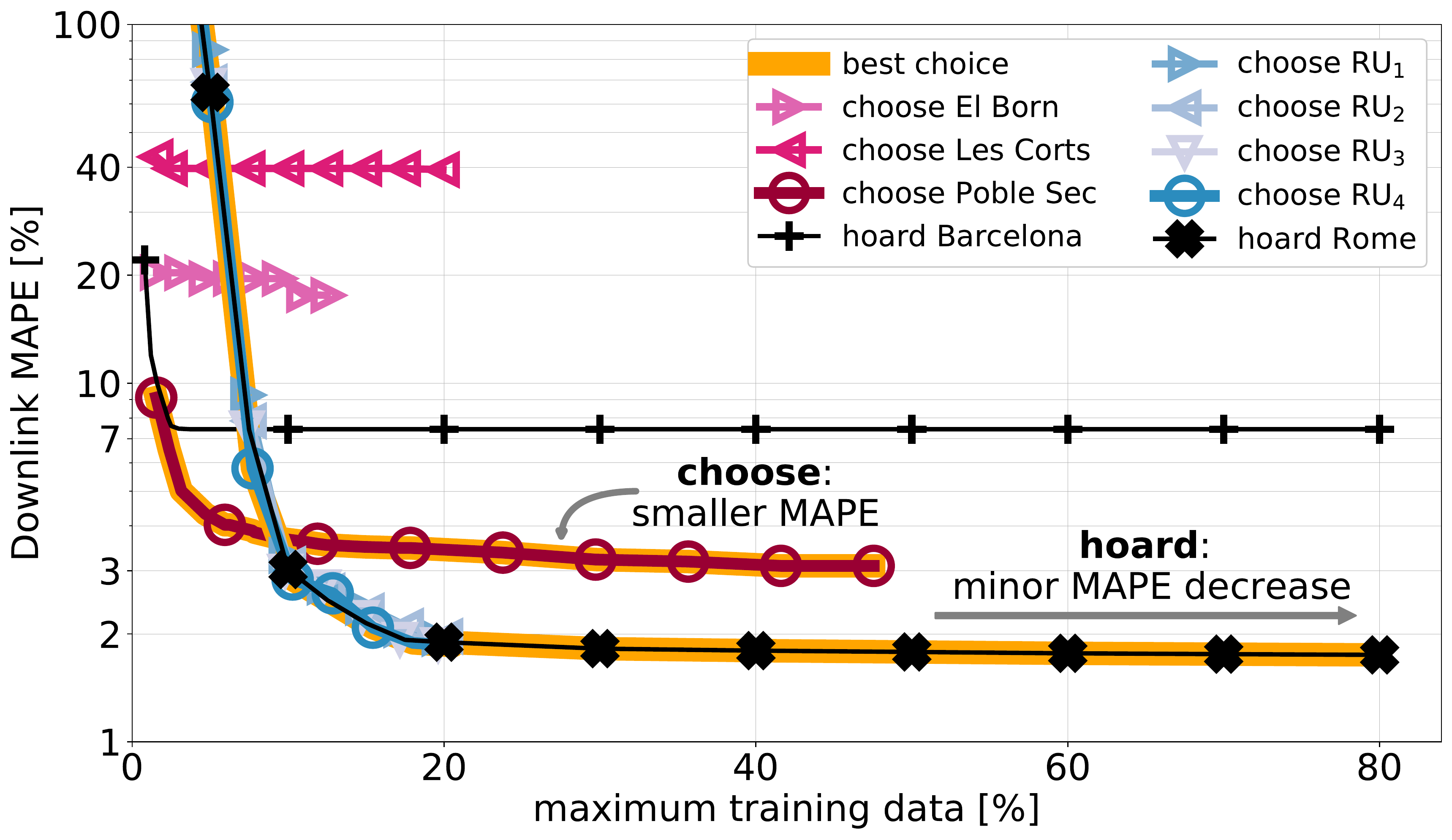}
    \caption{
    NN accuracy vs. quantity of training data
    in Rome/Barcelona (blue/purple), when the objective is to infer the performance of RU\textsubscript{4} (Rome) and Poble Sec (Barcelona).
    NN instances (lines) \emph{choose} training data from a
    RU (blue/purple lines), or \emph{hoard} data from all RUs (black).
    The best NN instance is also illustrated (orange line).
    }
    \label{fig:decrease-data}
\end{figure}

\Fig{decrease-data} shows how the DL MAPE (y-axis)
is impacted as we increase the amount of
training data (x-axis).
As mentioned above,
for the Rome (blue) and Barcelona (purple) datasets
we compare \emph{multiple} NN instances (lines) that
\emph{choose} training data from a single RU, and
an NN instance that \emph{hoards} data from all
RUs (black line).
Additionally,
\Fig{decrease-data} highlights what is the
best instance choice (thick, orange line) as we increase
the available data.

%

\Fig{decrease-data} shows that in Barcelona
the best MAPE (as low as 3.14\%) is always achieved by the
NN instance \emph{choosing} data from
the target RU at Poble Sec.
Owing to the heterogeneity of the scenario, when we \emph{hoard} training data to use at a single NN instance, the
MAPE never goes below
7.87\%, even using all the training
data.

In the Rome dataset, the MAPE at
RU\textsubscript{4} is essentially
the same for all NN instances, already using
less than 20\% of the data.
Specifically, with 20\% of the data,
an NN instance achieves 1.85\%
MAPE either if it
\emph{chooses} data from a single RU,
or \emph{hoards} data from all
Rome RUs; such an effect is due to the homogeneity of the Rome scenario. The {\em hoarding} strategy only reduces MAPE by a further 0.10\%, i.e.,
1210~bits per second in the URLLC scenario.

Overall, regardless of the amount of training data and their level of heterogeneity,
the benefit of \emph{hoarding} data from all RUs over \emph{choosing} is always limited -- and often there is no benefit at all. In some cases, the benefit is only evident when data is extremely rich; nonetheless it comes with an associated overhead, consistently with our intuition that not all data is always necessary.

\subsection{Impact of the maximum training time}
\label{sec:training-more-less-time}

\begin{figure}[t]
    \includegraphics[width=\columnwidth]{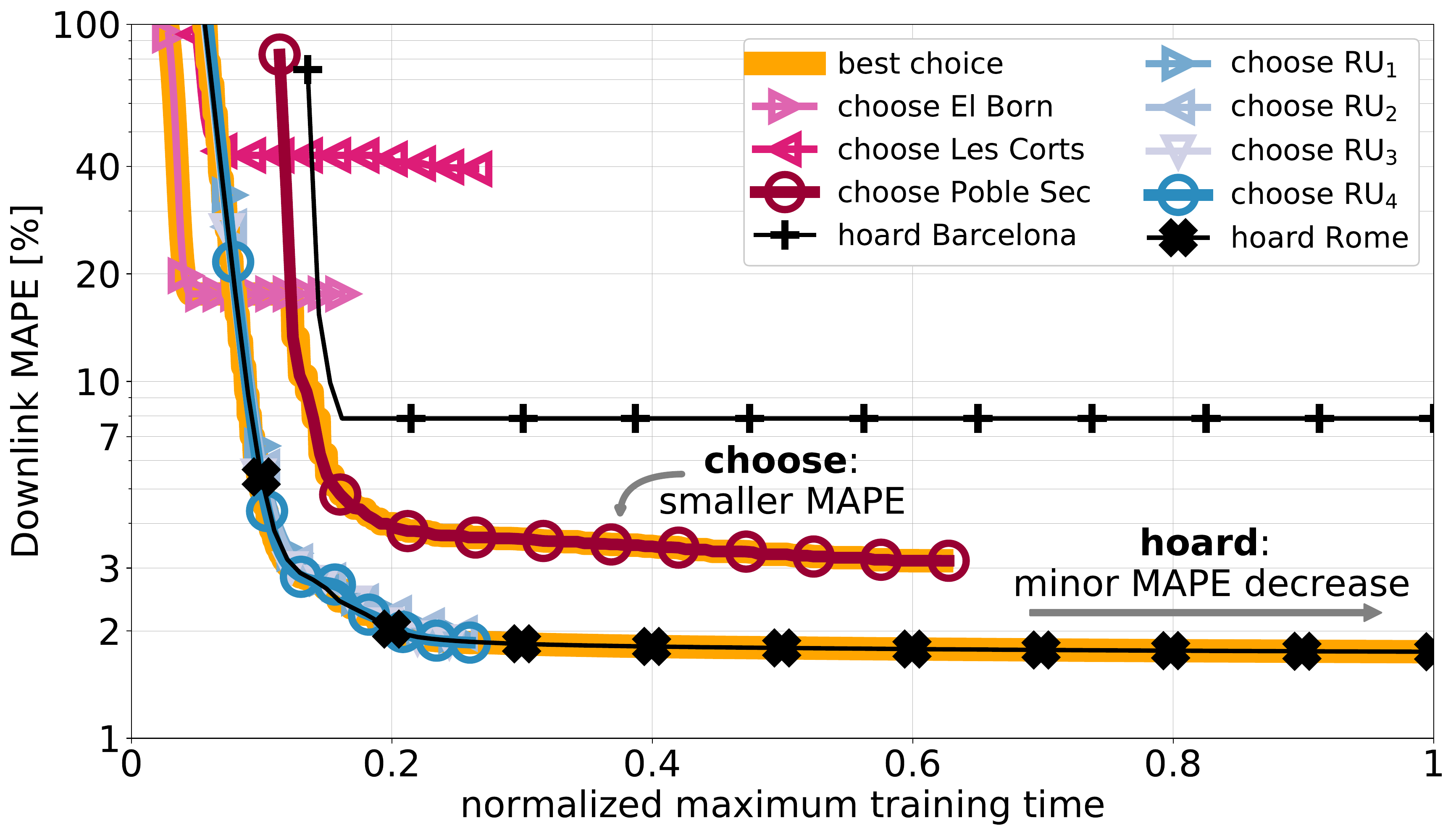}
    \caption{
    NN accuracy vs. training time
    in Rome/Barcelona (blue/purple), when the objective is to infer the performance of RU\textsubscript{4} (Rome) and Poble Sec (Barcelona).
    NN instances (lines) \emph{choose} training data from a
    RU (blue/purple lines), or \emph{hoard} data from all RUs (black).
    The best NN instance is also illustrated (orange line).
    }
    \label{fig:training-time-bs4}
\end{figure}

In \Fig{training-time-bs4}
we use all the training data from
Rome (blue) and Barcelona (purple),
and study the prediction error (y-axis) as
we increase the maximum training duration
(x-axis), 
normalized to how long it takes to run 100 epochs under the \emph{hoarding} strategy.

\Fig{training-time-bs4} shows that
the best MAPE in Barcelona is achieved
under the \emph{choosing} strategy -- from El Born for a normalized training time below 0.12, and from Poble Sec for longer training times. In the latter case, a remarkably low MAPE of 3.14\% is achieved. The MAPE of the {\em hoarding} strategy, in such a diverse scenario, cannot go below
7.87\%.

For the Rome dataset the MAPE at RU\textsubscript{4}
is essentially
the same if the NN instance is
trained for no longer than 0.25 time units, for all strategies; specifically, the MAPE is around 1.85\%. In this more homogeneous scenario, the {\em hoarding} strategy yields a minor 0.10\% performance improvement.

In general, independently of the maximum
training time and the diversity of the data,
it is again beneficial to \emph{choose}
training data from the target RU.
Furthermore, \Fig{training-time-bs4} 
does not report the network transfer delay,
which increases with more data and/or data from faraway locations~\cite{malandrino2021network}.
Such a delay is higher for the {\em hoarding} strategy, hence, considering it would further increase the attractiveness of the 
 \emph{choosing} strategy.

\subsection{Quality predictor xApp}
\label{sec:quality-predictor}

For concreteness, we focus on a quality predictor (QP) xApp that checks if an mMTC UE will have sufficient bandwidth when connecting to RU\textsubscript{4}. The QP xApp leverages the NN of~\Sec{training-more-less-time} which is fed with O-RAN data coming from RUs through the E2 interface, and provides near real-time estimations of the DL that UEs will experience.
These experiments are performed only with the Rome dataset, as the Barcelona one lacks the 
bitrate information.

\begin{figure}[t]
    \includegraphics[width=\columnwidth]{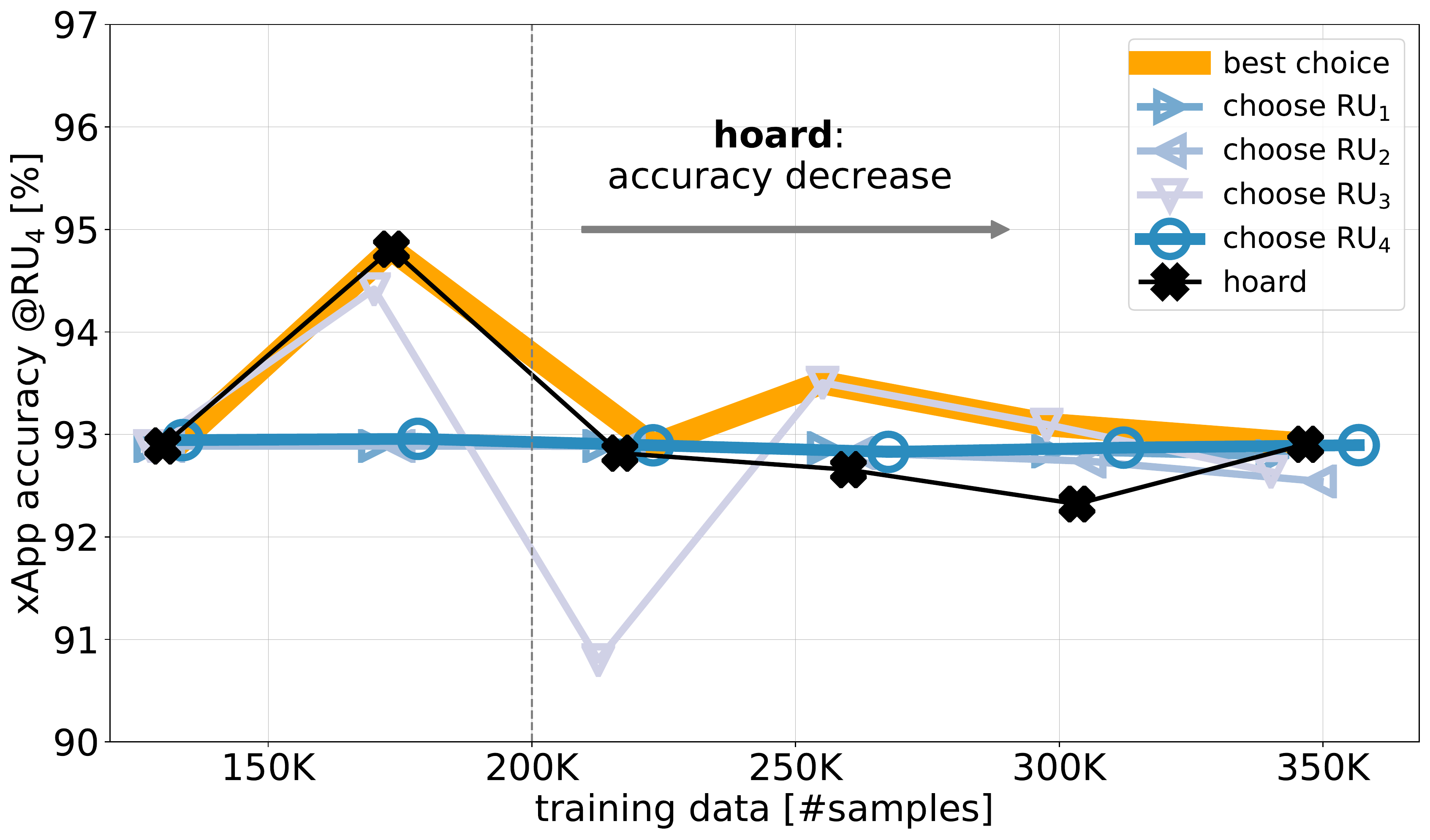}
    \caption{Accuracy of a QP xApp as a function of the quantity of training data. NN instances that \emph{choose} training data from
    one RU (blue lines) or \emph{hoard} training data
    from all RUs (black line) are used.}
    \label{fig:xapp}
\end{figure}

For our QP xApp the performance is expressed through the classification accuracy in assigning UEs seeking to connect to RU\textsubscript{4} to either the ``zero-bitrate'' or ``non-zero-bitrate'' classes.
\Fig{xapp} illustrates the xApp accuracy (y-axis)
when the quantity of training data increases (x-axis)
under the {\em choose} and {\em hoard} strategies.
%
It shows that it is better 
to use an NN instance that \emph{hoards}
training data from all the RUs.
As shown in \Fig{xapp}, the xApp achieves its best accuracy (namely,
94.81\%) under the {\em hoarding} strategy (specifically, when 175K data samples from all RUs are used). The {\em choosing} strategy results in a marginally smaller accuracy, namely, 93\%.

Interestingly, when there are between 250K and 300K samples, the best performance under the {\em choosing} strategy is obtained when the training data comes from RU\textsubscript{3}, i.e., not the same RU that will use the prediction.
Such a counter-intuitive behavior comes from the homogeneity of the dataset~\cite{bonati2021intelligence} and the large quantity of data available for RU\textsubscript{3}; in cases like this, the performance of a RU may indeed be best predicted through data from a {\em different} RU.

As in \Sec{training-more-less-time},
the xApp accuracy does not increase
after a certain amount
of training data; in our specific case, after 200K samples -- see~\Fig{xapp}.
For example, when the xApp uses the
NN instance that
\emph{hoards} training data from all RUs
(black line),
its accuracy drops down to 92.32\%
with 305K samples.
This highlights the relevance of
not only \emph{choosing} or \emph{hoarding}
the RUs used in the training stage,
but also deciding the fraction of data
used. 


The different levels of heterogeneity of the Rome and Barcelona datasets make them intuitively more adequate for different data selection strategies.
The {\em hoarding} strategy is naturally suited to homogeneous conditions like those of the Rome dataset; conversely, heterogeneous scenarios may benefit more from \emph{choosing} RU data, owing to the ability to create multiple model instances that fit the traffic of each RU. Interestingly, the \emph{choosing} strategy results in consistently good performance in both heterogeneous {\em and} homogeneous scenarios.
Widening our focus, we now discuss how our findings fit into a more generic problem of information-to-model matching for AI in O-RAN.

\section{Why choosing works:\\between networking and learning}
\label{sec:omatch}

The high-level ambition of this paper is to shift the focus from {\em how} to best combine all available data within one model instance to finding the best {\em matching} between data, model instances, and gNBs. Such a shift is motivated by three main factors, related to networking, ML, or both.

Our numerical results show that deviating from the standard approach of creating a single model instance using all the available data, can offer significant performance advantages; in other words, choosing {\em works} better than hoarding. We now switch our focus towards {\em why} choosing works, and remark how it helps to address three of the main issues affecting learning in O-RAN scenarios.

The first issue concerns the networking side of O-RAN and stems from the large cost of ML training. Whether such training is performed in a centralized or distributed manner, it always requires moving significant quantities of information -- data, gradients, models... -- around the network.
3GPP networks are divided into different planes (i.e., control, user and synchronization planes) that ensure the proper communication and management. 
Transferring large amounts of data may, thus, have an impact in all planes. Specifically, network saturation due to ML data transfers may significantly impact the transport in the synchronization plane by, e.g., increasing jitter due to queues full of data. 
Furthermore, the
ML training process will compete for bandwidth and computational resources in the user-plane, and potentially impact the performance of the latter. 
It follows that reducing the quantity of data that ML models are trained upon, if it can be done without jeopardizing the quality of the resulting decisions, is a very appealing prospect. 

The second issue is related to the ML side, more specifically, the relationship between the quantity of available data, the learning performance, and the training time. Theoretical and experimental results concur that the time taken by each training epoch grows {\em linearly} with the quantity of used data, while the learning quality improves more slowly, typically, according to a square-root law~\cite{malandrino2021network}. It follows that, while more data does translate into better learning, prolonged training times may not be worth it in time-constrained scenarios.
\begin{figure}[b]
    \centering
    \includegraphics[width=.48\columnwidth]{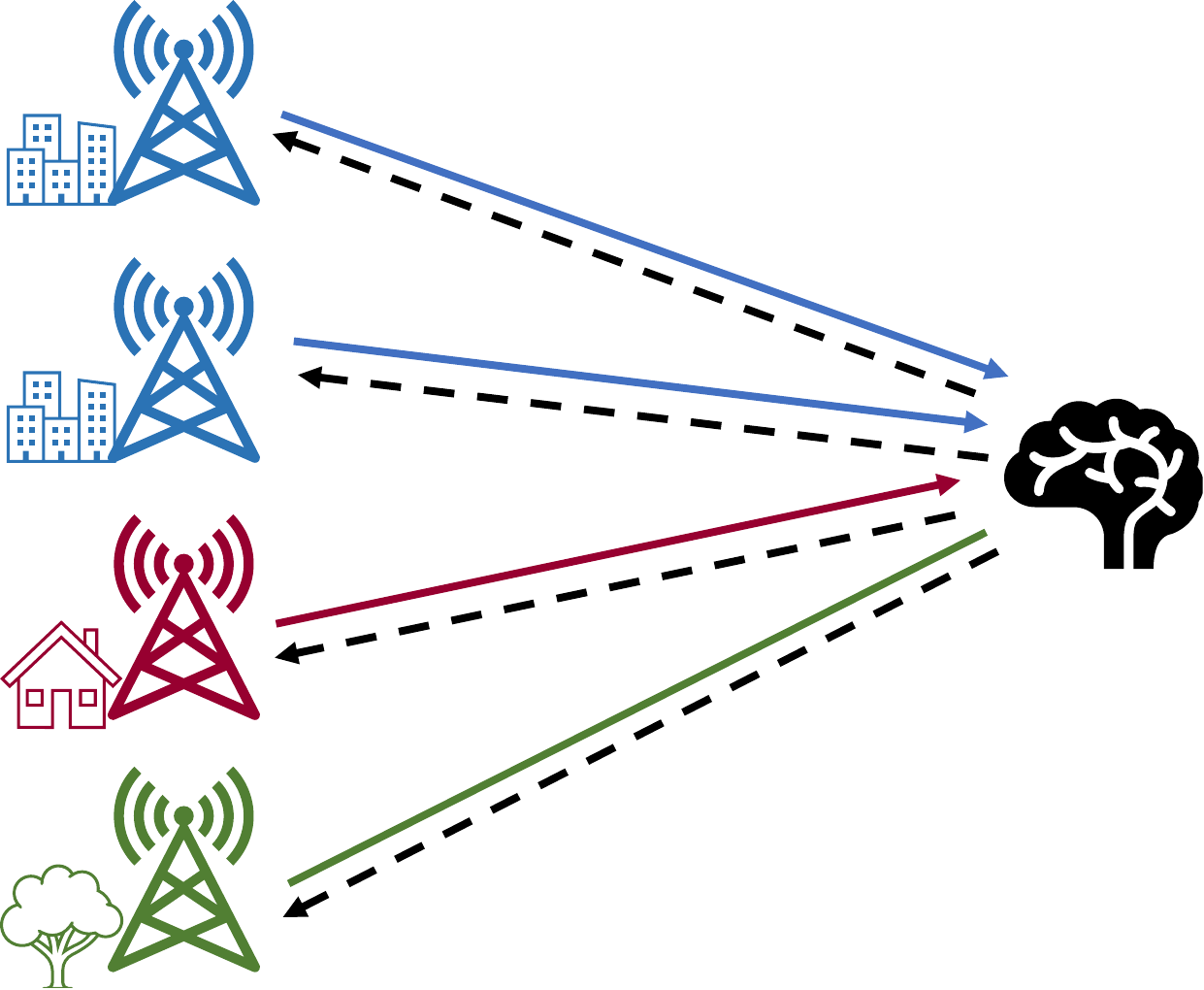}\hfill\includegraphics[width=.48\columnwidth]{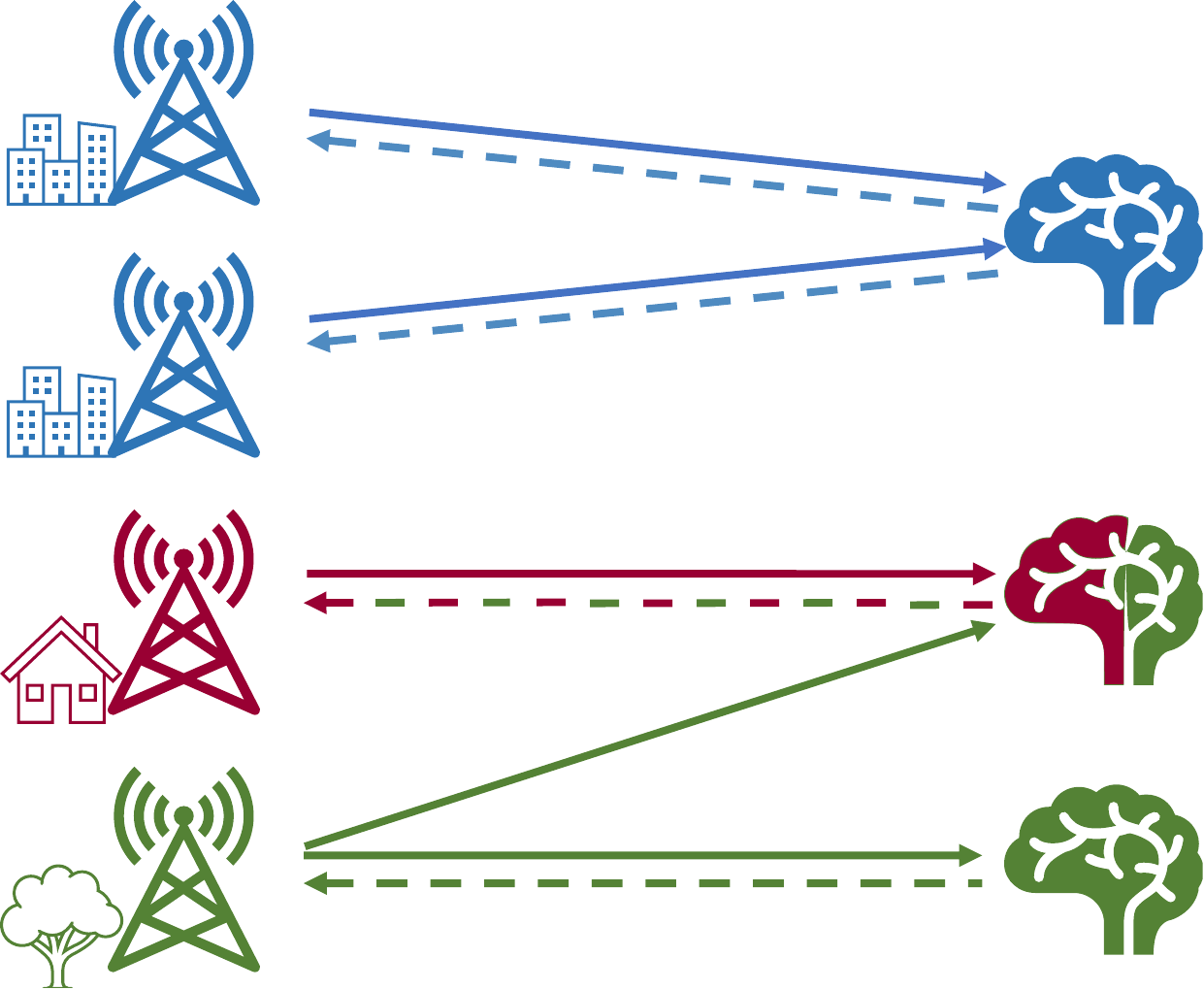}
    \caption{
        Scenario with urban (blue),
        residential (purple), and rural (green) locations.
        The traditional approach (left) \emph{hoards}
        all RUs' data (continuous lines) to a model
        instance taking decisions (dashed lines) at all
        locations.
        Our proposal (right) flexibly associates model instances, \emph{chosen}
        RUs to gather data, and locations to take decisions.
        \label{fig:balls}
    } 
\end{figure}

The networking and ML sides combine in the third issue, namely, the extent to which it is beneficial to learn from {\em heterogeneous} data. Indeed, different RUs may operate in very different conditions, e.g., rural/residential/urban areas, possibly with different traffic patterns and user mobility,
and even different technologies. Such heterogeneous conditions may result in heterogeneous {\em data} being fed to the ML model.
An ML model instance can learn from heterogeneous data, but that requires more complex models, which in turn have longer training times. The issue is so significant that some recent works on FL envision dropping nodes with overly-heterogeneous data from the training process.

Choosing -- more accurately, {\em being able} to choose -- addresses the concerns above in three main ways:
\begin{itemize}
    \item {\em multiple} model instances are allowed;
    \item each model instance can leverage information from some (or all) locations for its training;
    \item locations can use any model instance for their decisions, including those not trained using local information.
\end{itemize}
\Fig{balls}(right) represents a possible decision made where data is {\em chosen} and not {\em hoarded}: data from the two urban (blue) RUs is combined in one model instance, which both RUs then leverage. Data from the residential (red) and rural (green) RUs is kept separate and used for two different model instances; furthermore, the red+green model instance also uses data from the rural RU. Compared with \Fig{balls}(left), summarizing the state-of-the-art approach, we are creating more model instances, training each of them with a smaller quantity of data, and choosing how to {\em match} data and model instances.

Importantly, the one in \Fig{balls}(left) is {\em also} a possible decision. Indeed, being able to choose does not prevent falling back to creating a single model instance -- leveraging all information -- whenever the scenario and conditions warrant it.

\section{Open issues}
\label{sec:discussion}

The greater flexibility afforded by being able to choose model instances and data comes at a cost in terms of new decisions and additional factors to consider when making them; this, in turn, opens up new exciting avenues for future research.

A first, major topic is represented by the relationship between learning quality and network overhead. Traditionally, it is assumed that achieving a high learning quality requires more data, which entails more network overhead. However, our results suggest that, in many scenarios, it is possible to achieve both, i.e., to have a high learning quality with a limited quantity of data -- crucially, data that does not need to travel long distances across the network --, hence, with a limited network overhead.

This raises the issue of {\em what} makes certain datasets and scenarios more amenable to choosing or hoarding.
Consistently with our experiments, we can conjecture that {\em hoarding} works best in homogeneous scenarios, where gathering data from multiple sources helps training; conversely, heterogeneous scenarios might be better tackled by creating multiple model instances and {\em choosing} the data to train them. Being able to assess {\em a priori} whether the scenario at hand is better suited for choosing or hoarding -- e.g., by computing data-related metrics such as similarity -- would greatly help in choosing the right approach, hence, improve performance.

Finally, our results highlight how ML performance, i.e., learning accuracy, does not immediately or directly translate into application performance. This is observed by comparing \Fig{training-time-bs4} and \Fig{xapp}, where minor differences in the learning quality result in more significant differences in the application behavior. This calls for further attention on the fact that ML accuracy is not necessarily the best metric to evaluate the possible learning approaches, as AI/ML in O-RAN is usually a means not an end.
Hence, it raises
the need of a better modeling of the system as to provide a deeper analysis and develop more efficient solutions.

\section{Conclusion}
\label{sec:conclusion}

We have proposed and analyzed a new approach to the integration of AI in O-RAN scenarios, allowing
to assign different model instances to each gNB of the network, and independently {\em choose} the
data each instance is trained on.
Our approach deviates from the state of the art in that it does not seek to train one model instance for the whole network and to train it using all available data; therefore, it provides
more flexibility than fully-centralized and fully-distributed approaches.

Our performance evaluation, leveraging real-world traces, shows how our approach yields very attractive trade-offs between training time and learning effectiveness, by combining data from different sources in a flexible manner. Future research directions stemming from our work include characterizing {\em a priori} the usefulness of data for AI training, trade-offs between
data transfer delays and AI training time, and the impact of AI accuracy over the performance of concrete applications.

\section*{Acknowledgment}

This work was partially supported by the European Union’s Horizon 2020 research and innovation program through the project iNGENIOUS under grant agreement No. 957216,
and by the Spanish Ministry of Economic Affairs and Digital
Transformation and the European Union-NextGenerationEU through the
UNICO 5G I+D 6G-EDGEDT and 6G-DATADRIVEN projects. 

\balance
\bibliographystyle{IEEEtran}
\bibliography{refs}%

\begin{thebibliography}{10}
\providecommand{\url}[1]{#1}
\csname url@samestyle\endcsname
\providecommand{\newblock}{\relax}
\providecommand{\bibinfo}[2]{#2}
\providecommand{\BIBentrySTDinterwordspacing}{\spaceskip=0pt\relax}
\providecommand{\BIBentryALTinterwordstretchfactor}{4}
\providecommand{\BIBentryALTinterwordspacing}{\spaceskip=\fontdimen2\font plus
\BIBentryALTinterwordstretchfactor\fontdimen3\font minus
  \fontdimen4\font\relax}
\providecommand{\BIBforeignlanguage}[2]{{%
\expandafter\ifx\csname l@#1\endcsname\relax
\typeout{** WARNING: IEEEtran.bst: No hyphenation pattern has been}%
\typeout{** loaded for the language `#1'. Using the pattern for}%
\typeout{** the default language instead.}%
\else
\language=\csname l@#1\endcsname
\fi
#2}}
\providecommand{\BIBdecl}{\relax}
\BIBdecl

\bibitem{garcia2021ran}
A.~Garcia-Saavedra and X.~Costa-Perez, ``{O-RAN: Disrupting the virtualized RAN
  ecosystem},'' \emph{IEEE Communications Standards Magazine}, 2021.

\bibitem{bonati2021intelligence}
L.~Bonati, S.~D'Oro, M.~Polese, S.~Basagni, and T.~Melodia, ``{Intelligence and
  learning in O-RAN for data-driven NextG cellular networks},'' \emph{IEEE
  Communications Magazine}, 2021.

\bibitem{ORAN_WG2_AI_ML}
{O-RAN Alliance}, ``{O-RAN Working Group 2. AI/MLWorkflow Description and
  Requirements (O-RAN.WG2.AIML-v01.02)},'' in \emph{Tech. Rep.}, 2021.

\bibitem{malandrino2021network}
F.~Malandrino, C.~F. Chiasserini, N.~Molner, and A.~De~La~Oliva, ``{Network
  Support for High-performance Distributed Machine Learning},'' \emph{IEEE/ACM
  Transactions on Networking}, 2022.

\bibitem{shafin2020artificial}
R.~Shafin, L.~Liu, V.~Chandrasekhar, H.~Chen, J.~Reed, and J.~C. Zhang,
  ``{Artificial intelligence-enabled cellular networks: A critical path to
  beyond-5G and 6G},'' \emph{IEEE Wireless Communications}, 2020.

\bibitem{pamuklu2021reinforcement}
T.~Pamuklu, M.~Erol-Kantarci, and C.~Ersoy, ``{Reinforcement learning based
  dynamic function splitting in disaggregated green Open RANs},'' in \emph{IEEE
  ICC}, 2021.

\bibitem{ayala2019vrain}
J.~A. Ayala-Romero, A.~Garcia-Saavedra, M.~Gramaglia, X.~Costa-Perez,
  A.~Banchs, and J.~J. Alcaraz, ``{vrAIn: A deep learning approach tailoring
  computing and radio resources in virtualized RANs},'' in \emph{ACM MobiCom},
  2019.

\bibitem{cao2021user}
Y.~Cao, S.-Y. Lien, Y.-C. Liang, K.-C. Chen, and X.~Shen, ``{User Access
  Control in Open Radio Access Networks: A Federated Deep Reinforcement
  Learning Approach},'' \emph{IEEE Transactions on Wireless Communications},
  2021.

\bibitem{nikbakht2020unsupervised}
R.~Nikbakht, A.~Jonsson, and A.~Lozano, ``{Unsupervised learning for C-RAN
  power control and power allocation},'' \emph{IEEE Communications Letters},
  2020.

\bibitem{pradhan2020computation}
C.~Pradhan, A.~Li, C.~She, Y.~Li, and B.~Vucetic, ``{Computation offloading for
  IoT in C-RAN: Optimization and deep learning},'' \emph{IEEE Transactions on
  Communications}, 2020.

\bibitem{bonati2020cellos}
L.~Bonati, S.~D’Oro, L.~Bertizzolo, E.~Demirors, Z.~Guan, S.~Basagni, and
  T.~Melodia, ``{CellOS: Zero-touch softwarized open cellular networks},''
  \emph{Computer Networks}, 2020.

\bibitem{MLedge_TMC2021}
M.~Polese, R.~Jana, V.~Kounev, K.~Zhang, S.~Deb, and M.~Zorzi, ``{Machine
  Learning at the Edge: A Data-Driven Architecture With Applications to 5G
  Cellular Networks},'' \emph{IEEE Transactions on Mobile Computing}, 2021.

\bibitem{kazemifard2021minimum}
N.~Kazemifard and V.~Shah-Mansouri, ``{Minimum delay function placement and
  resource allocation for Open RAN (O-RAN) 5G networks},'' \emph{Computer
  Networks}, 2021.

\bibitem{bashir2019optimal}
A.~K. Bashir, R.~Arul, S.~Basheer, G.~Raja, R.~Jayaraman, and N.~M.~F. Qureshi,
  ``{An optimal multitier resource allocation of cloud RAN in 5G using machine
  learning},'' \emph{Transactions on emerging telecommunications technologies},
  2019.

\bibitem{wang2021network}
S.~Wang, Y.~Ruan, Y.~Tu, S.~Wagle, C.~G. Brinton, and C.~Joe-Wong,
  ``{Network-aware optimization of distributed learning for fog computing},''
  \emph{IEEE/ACM Transactions on Networking}, 2021.

\end{thebibliography}
\begin{IEEEbiographynophoto} 
{Jorge Martín Pérez} got his Ph.D. in Telematics Engineering at Universidad Carlos III de Madrid in 2021, where he works as postdoc. 
\end{IEEEbiographynophoto}
\vskip -2.5\baselineskip plus -7fil
\begin{IEEEbiographynophoto} 
{Nuria Molner} obtained her Ph.D. from Universidad Carlos III de Madrid in 2021. Currently, she is a researcher at Universitat Polit\`ecnica de Val\`encia (iTEAM-UPV).
\end{IEEEbiographynophoto}
\vskip -2.5\baselineskip plus -7fil
\begin{IEEEbiographynophoto} 
{Francesco Malandrino} (M'09, SM'19) earned his Ph.D. degree from Politecnico di Torino in 2012 and is now a researcher at the National Research Council of Italy (CNR-IEIIT). 
\end{IEEEbiographynophoto}
\vskip -2.5\baselineskip plus -7fil
\begin{IEEEbiographynophoto} 
{Antonio de la Oliva} got his M.Sc. in 2004 and his Ph.D. in 2008, is an associate professor at Universidad Carlos III de Madrid.

\end{IEEEbiographynophoto}
\vskip -2.5\baselineskip plus -7fil
\begin{IEEEbiographynophoto} 
{Carlos J. Bernardos} got his M.Sc. in 2003, and his Ph.D. in 2006, is an associate professor at Universidad Carlos III de Madrid.
\end{IEEEbiographynophoto}
\vskip -2.5\baselineskip plus -7fil

\begin{IEEEbiographynophoto} 
{David Gomez-Barquero} is an associate professor at Universitat Polit\`ecnica de Val\`encia and subdirector of research at iTEAM Research Institute. 
\end{IEEEbiographynophoto}

\end{document}